# Study of discrete boundary layer suction for transition delay

Mukund R. [1,a], Pranav Joshi[1,2,b], & Ishan Singh[3,c]

[1] Formerly at Experimental Aerodynamics Division, CSIR-National Aerospace Laboratories, Bengaluru 560017, Karnataka, India
[2] Present address: Department of Mechanical Engineering, Indian Institute of Technology Kanpur, Kanpur 208016, Uttar Pradesh, India
[3] Experimental Aerodynamics Division, CSIR-National Aerospace Laboratories, Bengaluru 560017, Karnataka, India

[a] mukund13@yahoo.com
[b] jpranavr@iitk.ac.in, ORCID ID: 0000-0001-7368-0607
[c] ishan@nal.res.in, ORCID ID: 0000-0002-2603-6496

## Abstract

In the present study we have assessed the effect of suction through discrete perforations on the transition of the Blasius boundary layer. In particular, the intermittency and the spectra of the streamwise velocity fluctuations, obtained using hot-wire anemometry, were used to evaluate the 'optimum' suction, i.e., one that provides the greatest delay in transition. The displacement thickness Reynolds number at suction was varied between 1400 and 1600, while the number of rows of holes was either four or ten. The suction holes had diameter of the order of the displacement thickness of the boundary layer, a practical size from manufacturing and maintenance perspectives. As the suction strength was increased the transition front moved downstream, with the maximum delay in transition occurring at the 'optimum' suction. When the suction was increased further, the transition front moved upstream again. The results, supported by a scaling analysis, showed that suction volume flow rate is a better indicator of the effects of suction on the transition delay than suction velocity. In particular, suction through different numbers of rows of perforations produces largely similar effects on the boundary layer transition, provided the suction volume flow rate remains the same. However, suction through fewer rows provided a slightly greater delay in transition in addition to a slightly lower optimum suction volume flow rate. The results also showed that suction applied at a lower Reynolds number provides a substantially greater transition delay but at slightly higher values of the normalized optimum suction volume flow rate.

# 1 Introduction

Suction through the wall of a laminar boundary layer can delay its transition to turbulent state. Such laminar flow control helps in the eventual goal of drag reduction (Joslin 1998). Uniformly distributed suction delays the transition of the boundary layer as it decreases the local Reynolds number of the boundary layer and also makes the velocity profile fuller, which is more stable to perturbations (Schlichting and Gersten 2000). Suction through a porous wall comes closest to ideal uniform suction and its stabilizing effect on the Tollmien-Schlichting waves has been studied both experimentally (Reynolds and Saric 1986) and numerically (Reed and Nayfeh 1986; Huang and Wu 2016). Though suction using porous strips and slots has been widely studied, they have not been implemented commercially due to practical considerations (Gregory 1961; Saric 1985) and have been superseded by uniformly distributed perforations, as used in the present study.

The flow downstream of an isolated perforation with suction is three-dimensional with features such as counter-rotating streamwise vortices and inflection points in the velocity profiles (Gregory 1961; Saric 1985; Meitz and Fasel 1995; MacManus and Eaton 1998; Sattarzadeh and Fransson 2017). For a low suction velocity, $v_s$, these streamwise vortices and the associated low- and high-speed streaks can stabilize the Tollmien-Schilichting (TS) waves and delay transition (Sattarzadeh and Fransson 2017). However, if the suction velocity is increased beyond the 'optimum' (to be defined shortly) value, these flow features can destabilize the flow and advance the transition front (MacManus and Eaton 2000). As the suction velocity is increased further to what is known as the 'critical' suction velocity, $v_{s,c}$ (Gregory 1961), the transition front moves back to the same location as that without suction.

Much of the prior work has focused on the dependence of the critical suction velocity on the suction parameters. For an isolated perforation, irrespective of the Reynolds number $Re_{\delta^*} = U_\infty \delta^*/\nu$ at suction, the normalized critical suction velocity $v_{s,c}/U_\infty$ increases as $d/\delta^*$ decreases (MacManus and Eaton 2000), becoming 'practically unlimited' for $d/\delta^* < 0.6$ (MacManus and Eaton 2000). Here, $U_\infty$ is the freestream velocity, $d$ is the perforation diameter, $\delta^*$ is the displacement thickness of the boundary layer, and $\nu$ is the kinematic viscosity of the fluid. For a single row of perforations, $v_{s,c}/U_\infty$ increases as $d/\delta^*$ decreases at a constant $Re_{\delta^*}$, and decreases as $Re_{\delta^*}$ at suction increases (Gregory 1961). The critical suction velocity attains a minimum for a normalized spanwise pitch of $2< p/d <5$ (Gregory and Walker 1958) indicating significant inter-hole effects. However, $v_{s,c}/U_\infty$ increases as $p/d$ is increased beyond 5 and plateaus for $p/d \gtrsim 10$ (Butler 1957), suggesting that the inter-hole effects are negligible for large $p/d$. For more than three rows of perforations, the critical suction velocity has been found to be 'practically unlimited' for very small spanwise pitch $p_s/d < 3$ and small perforation diameter $d/\delta^* < 2$ (Gregory and Walker 1958). Some researchers have reported an increase (Gregory and Walker 1958) while others have observed a decrease (Goldsmith 1957; Blanchard et al. 1991) in the critical suction velocity with an increase in the number of rows, $n$.

Few studies have been performed on the details of the flow structure downstream of the suction holes. It has been shown that the streamwise vortices emanating from closely-spaced adjacent holes in a row can merge to form horseshoe vortices at a moderately high suction strength (Gregory 1962). Furthermore, the presence of downstream suction holes can change the morphology and increase the strength of the streamwise vortices emanating from a hole (MacManus and Eaton 1996). The experiments of Becker and Jovanovic (2010) with suction through a single row of perforations showed that the turbulence intensity profiles behind a suction hole showed a disturbance peak, which got damped out with downstream distance for subcritical suction. In the case of supercritical suction, it got amplified, eventually leading to an early transition of the boundary layer.

Wright and Nelson (2001) used four closely-spaced porous suction strips having randomly distributed 100 micron diameter perforations on a flat-plate-aerofoil. They used an algorithm to optimize suction through each of the four strips to extend the laminar flow while minimizing the pumping energy. Crowley and Atkin (2016, 2018) performed experiments with discrete perforations ($p/d$ =10, $d/\delta^* \approx 0.5$) and observed that medium suction suppresses natural and excited (by loudspeaker) disturbances but introduces its own low-frequency disturbances. In their second paper (Crowley and Atkin 2018), they showed that excessive suction leads to transition through a process involving amplification of the high-frequency broadband velocity fluctuations, accompanied by an inflection point in the velocity profile. Using perforated suction panels with $p/d$ =10 and $d/\delta^* \sim O(0.1)$, Grappadelli et al (2021) observed that as suction was increased the transition location initially moved downstream rapidly, and then gradually.

An ideal uniform suction would be amenable to theoretical analysis which renders an easier design process. Effects similar to those of uniform suction were achieved by Saric and Reed (1986) using a very small hole diameter, $d/\delta^* =$ 0.026, and a moderately large spanwise pitch, $p/d$ =10. However, achieving $d/\delta^* \sim O(< 0.1)$ seems difficult in practical applications such as aircraft wings where $\delta^*$ would be in the range of 100-500 $\mu$m. Schrauf and von Geyr (2020) used modern laser drilling techniques to achieve hole diameters of 55 $\mu$m and expressed difficulties in the drilling process not to mention the maintenance issues. Practically, a minimum hole diameter of 200-500 $\mu$m could be achieved, for which $d/\delta^* \sim O(1)$. Suction through holes with such moderately high $d/\delta^*$ cannot be treated as uniform ideal suction for theoretical analysis. More importantly, the destabilizing effects of strong suction are important for high $d/\delta^*$, as evidenced by the existence of a critical suction velocity. These issues underline the need for studies to understand the effects of discrete suction using $d/\delta^* \sim O(1)$ on boundary layer stability and the transition process, and to develop better models for transition prediction.

While prior research has focused on critical suction, the value of optimum suction, defined here for a given flow as the suction strength (flow rate or velocity) that provides the longest delay in the onset of transition, is more important for designing efficient suction systems. Bokser et al (1997) showed the existence of an optimum suction strength by measuring the drag of a wing. Some studies, e.g., Gregory and Walker (1958) and Butler (1957), also reported the variation of the transition location of the boundary layer with suction velocity on airfoils. However, the maximum extension of the laminar flow in these studies was limited by the adverse streamwise pressure gradient on the aft portion of the airfoils, thus precluding the measurement of the optimum suction strength.

The present work aims to address this gap in our knowledge by performing experiments to study the effects of suction through perforations with $d/\delta^* \sim O(1)$ on the Blasius boundary layer. Optimum suction velocity and volume flow rate are measured for two suction configurations that differ in the number of rows of holes, maintaning $Re_{\delta^*}$ at suction between 1400 and 1600. We show that suction volume flow rate is a better indicator of the suction strength than the suction velocity and that applying suction through shorter suction strips and at a lower Reynolds number is, in general, advisable. The database generated should prove useful to validate methods for the prediction of the effects of discrete suction and also serve as the basis for further systematic studies of the impact of surface curvature and streamwise pressure gradient on boundary layers subjected to suction.

## 2 Experimental setup and methods

### 2.1 Experimental setup

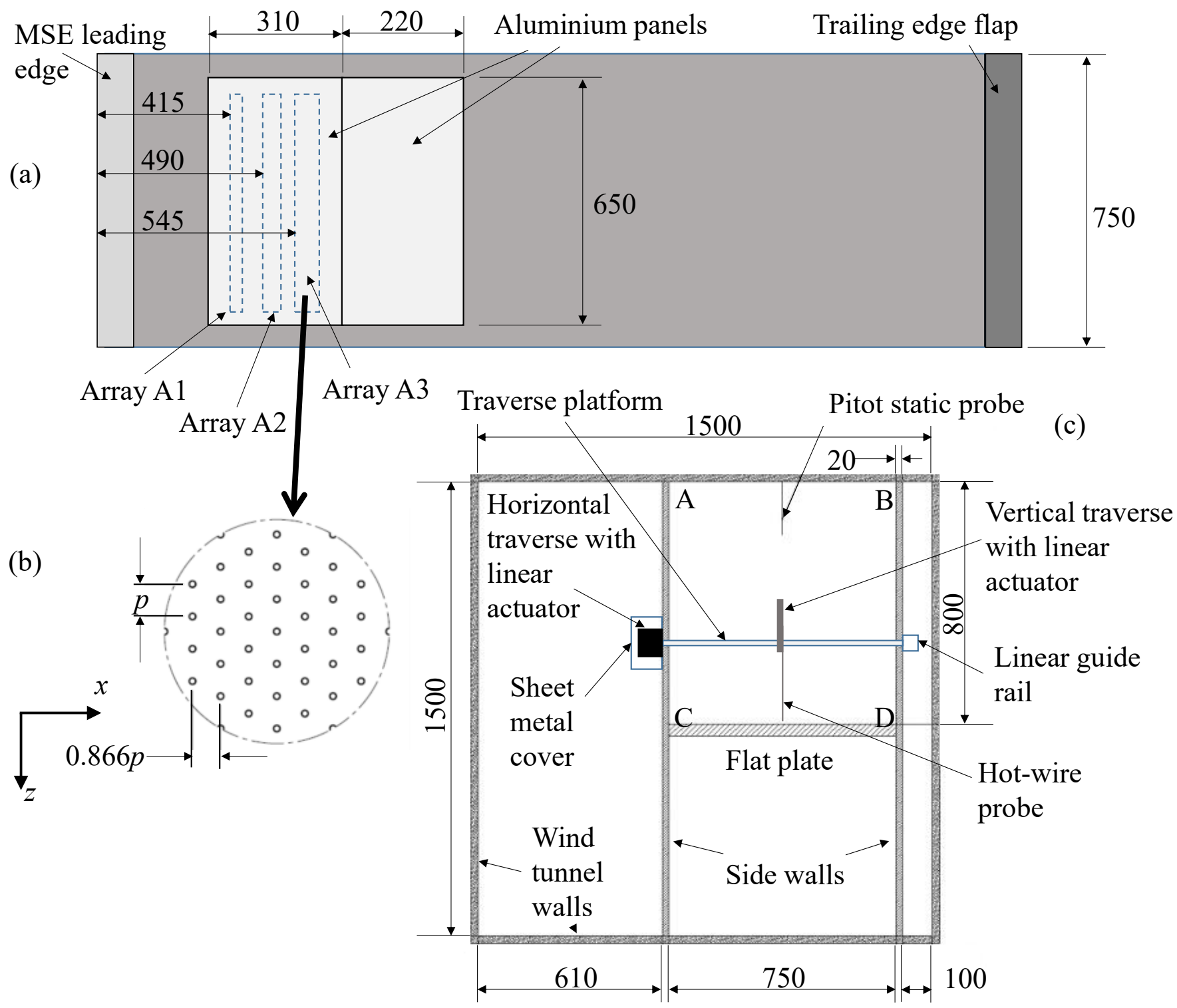

**Fig. 1 a)** The boundary layer plate and the aluminum panels, **b)** enlarged view of the arrangement of suction perforations in an array, **c)** mounting details of the plate and the traversing mechanism. All dimensions are in mm. Drawings are not to scale

The experiments were conducted in the 1.5 m × 1.5 m low-speed wind tunnel in the Experimental Aerodynamics Division of the National Aerospace Laboratories (CSIR-NAL). The freestream velocity in the tunnel could be controlled to within 1% and the freestream turbulence level was in the range of 0.09% to 0.1% for the range of velocities used in the present study. A 2.3 m long, 0.75 m wide, and 12 mm thick flat plate (Fig. 1 a) was moulded out of glass-fibre reinforced plastic (FRP) with an embedded aluminium framework used for strengthening and mounting the plate. The flat plate was mounted between two vertical sidewalls installed in the test section as sketched in Fig. 1 c. The $x$-, $y$- and $z$-axes of the coordiante system were along the streamwise, wall-normal and spanwise directions, respectively, with the origin located at the spanwise centre of the leading edge of the plate. The 125 mm long leading edge section of the flat plate was made of aluminium and had a modified super-elliptic cross-section, the contour of which was as recommended by Hanson et al (2012). The plate had two rectangular cut-outs as sketched in Fig. 1 a, into which two aluminium suction panels were mounted flush with the plate.

Three arrays of suction holes designated A1, A2, and A3 were drilled in the central region of the first suction panel as shown in Fig. 1 a, whereas no holes were drilled in the second panel. The arrays had 10 rows of suction holes each, and had spanwise pitches of $p/d$ = 5, 7, and 10, respectively. Successive rows of holes in each array were staggered in the spanwise direction by $p/2$ (see Fig. 1 b) and the streamwise separation between successive rows was $0.866p$.

On the one hand, the smallest pitch of $p/d = 5$ was chosen to avoid very low pitch values which may lead to the amalgamation of vortices emanating from adjacent holes (Gregory 1962). On the other hand, the largest pitch $p/d = 10$ was assumed to be large enough to make the inter-hole interactions negligible. The diameter of the perforation holes was increased slightly from 0.7 mm for array A1 to 0.8 mm for A2 and A3 to retain similar $d/\delta^*$ for all $p/d$. Any arrays or rows of holes could be used for suction by closing other arrays and rows from underneath using thin adhesive tapes. All the data presented in this paper used suction through the array A1 in which the suction holes extended in the streamwise direction from $x = 415$ mm to $x = 442$ mm.

A 15 mm deep rectangular tray made from FRP was attached to the underside of the suction panel to form a plenum chamber as shown in fig. 2 a. The plenum chamber was provided with three rows of static pressure ports for ascertaining the spanwise uniformity of suction pressure inside the chamber. One set of eight suction ports, 10 mm diameter each, was provided in the front (upstream) of the plenum chamber, and another identical set in the aft. For improving the flow uniformity one would choose the set farther away from the particular array of suction holes being used, closing the other set using adhesive tapes. Those set of ports used was connected to a manifold through 10 mm diameter flexible tubes of equal lengths. As shown in Ffig. 2 b, the manifold was connected through a flow meter (Testo 0555 6441), pressure control valves, and a bypass inlet to a 3 hp regenerative blower, which acted as a suction pump. For low suction flow rates, the bypass valve was kept fully open and the suction flow rate was controlled by opening/closing the in-line valve. For higher flow rates, the bypass valve had to be partially closed.

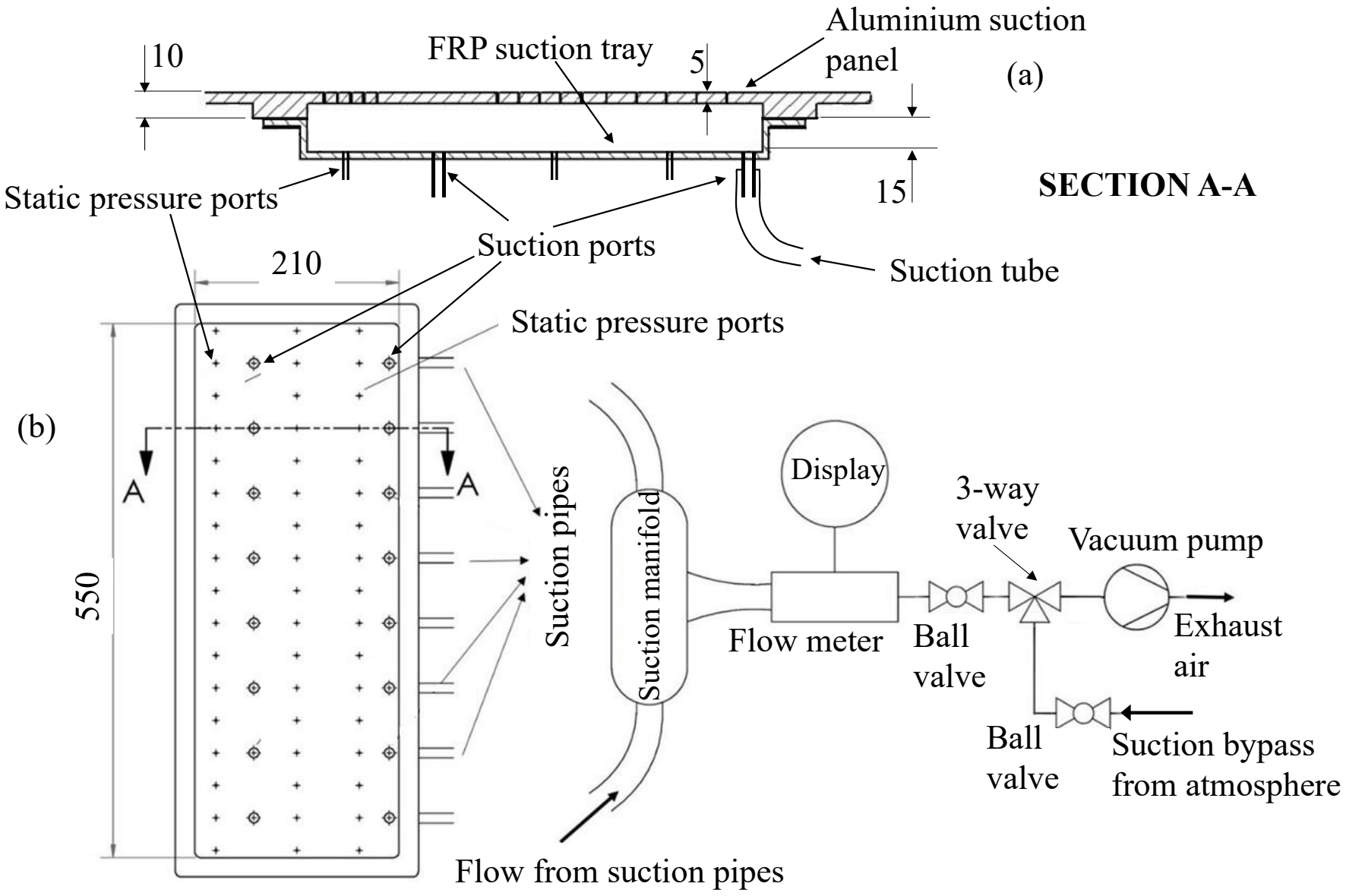

**Fig. 2** Schematic showing, **a)** the details of the suction tray and **b)** the suction system. All dimensions are in mm. Drawing is not to scale

The boundary layer velocity over the flat plate was measured using hot-wire anemometry. The hot-wire probe used for the purpose was traversed vertically in the $z = 0$ plane by using a linear actuator (THK KR20) attached to a streamlined aluminium platform positioned 235 mm above the plate, as shown in Fig. 1 c. The platform spanned the width of the flat plate and extended beyond the two side walls through streamwise slots provided in the side walls.

The platform was attached to a horizontal linear actuator (THK GL15N) on one side and a linear guide rail and block (THK HSR20) on the other side. The two actuators were connected to respective stepper motors, which were controlled from the computer. The whole assembly enabled smooth traversing of the probe over streamwise ($x$) and wall-normal ($z$) distances of 1354 mm and 80 mm with resolutions of 0.5 mm and 30 μm, respectively.

## 2.2 Instrumentation

A pitot-static tube was inserted through the top wall above the flat plate (Fig. 1 c) approximately 300 mm downstream of the leading edge and connected to a 200 mm $H_2O$ pressure gauge from Furness Corp to obtain the freestream velocity above the flat plate. Static pressure ports located on top of the flat plate and those located inside the plenum chamber were used to measure the respective pressures (when required) using 70 mm ESP scanners from Pressure Systems. The output from the flow meter was measured using a Masibus 1008S flow indicator-totalizer, which displayed the volume flow rate and also provided the corresponding voltage output for the data acquisition card. Velocity measurements were performed in the boundary layer using a Dantec 55P11 miniature single-wire probe connected to an AN1002 constant temperature anemometer from A.A. Lab Systems Ltd. A LabVIEW program was written to traverse the probe to the required position and to acquire data simultaneously from the hot-wire anemometer, the mass flow meter, and the pressure gauges using a National Instruments-based data acquisition system. The whole instrumentation setup including the anemometer and the data acquisition computer were electrically isolated by sourcing power through a 5 kW UPS that was grounded independently.

## 2.3 Preliminary measurements

The streamwise variation of the static pressure along the flat plate was measured using pressure ports located all along the centre-line of the FRP section on the flat plate. Initial measurements showed a small streamwise pressure gradient on the plate caused by the blockage due to the suction piping and manifold located below the flat plate. This was corrected by a combination of adjusting the pitch angle of a flap located at the trailing edge of the flat plate and creating a small blockage in the test section using a course cotton mesh attached at the rear above the plate (at location marked ABCD in Fig. 1 c). The resulting streamwise variation of the pressure coefficient $C_p = (p - p_\infty)/(0.5\rho U_\infty^2)$ on the flat plate is shown in Fig. 3 a for $U_\infty = 30$ m/s and three streamwise positions of the traverse. Here, $p_\infty$ is the reference static pressure measured by the pitot static tube above the flat plate and $\rho$ is the density of the fluid. It is evident that the static pressure varied negligibly along the length of the plate. The variation of the static pressure when the traverse was located at $x = 1.3$ m (black line in Fig. 3 a) shows that the effect of the traverse on the $C_p$ distribution extends approximately 0.2 m upstream of the traverse. Consequently, the hot-wire probe was maintained at a streamwise distance of 0.25 m from the traverse, thus avoiding the effect of the traverse mechanism on the measurements.

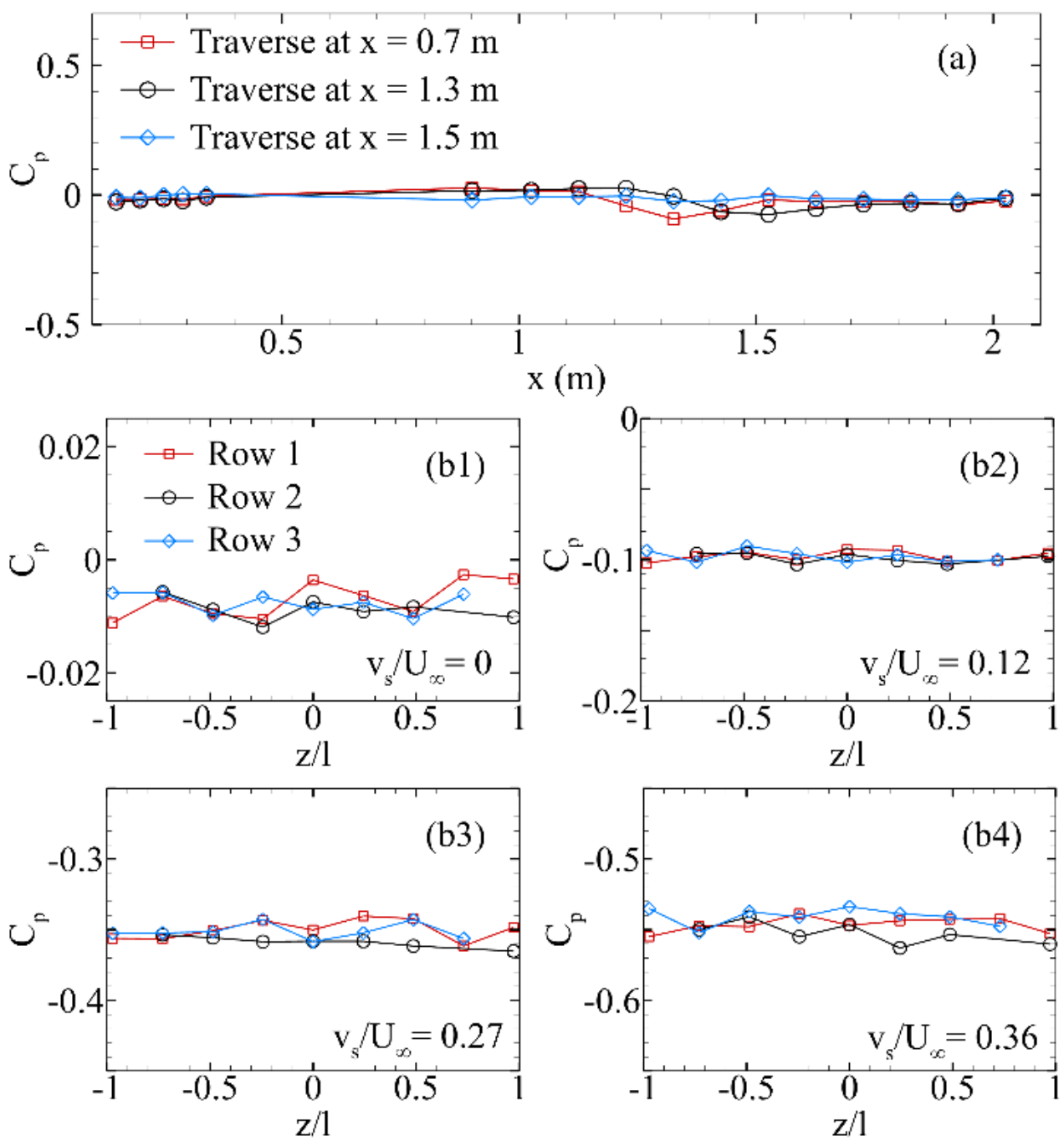

**Fig. 3 a** Variation of the coefficient of static pressure in the streamwise direction on the flat plate at $U_\infty = 30$ m/s for different streamwise positions of the leading edge of the traverse, **b1** to **b4** show the spanwise variation of the coefficient of static pressure in the plenum chamber for $U_\infty = 30$ m/s and suction through the array A1 (all rows open, i.e., $n = 10$) for various strengths of suction. Here, $l = 275$ mm is the half-span of the suction tray

The static pressure inside the plenum chamber was measured using the three rows of pressure ports provided in the floor of the chamber. The $C_p$ results for $U_\infty = 30$ m/s and suction at different rates through the first array (A1) of holes are shown in Fig. 3 b. The spanwise variations in $C_p$ are within ±0.004 for the flow without suction and increase to ±0.01 for the highest suction rate. These small values indicate a reasonably uniform suction.

The boundary layer mean velocity profiles at $x = 0.45$ m and $x = 1.4$ m without suction are shown in fig. 4 a, and the amplitude spectra for the velocity fluctuations at $x = 1.4$ m and $y = 3$ mm ($y/\delta \approx 0.2$) in fig. 4 b. The laminar boundary layer at $x = 0.45$ m compares well with the Blasius boundary layer, while that at $x = 1.4$ m is fully turbulent.

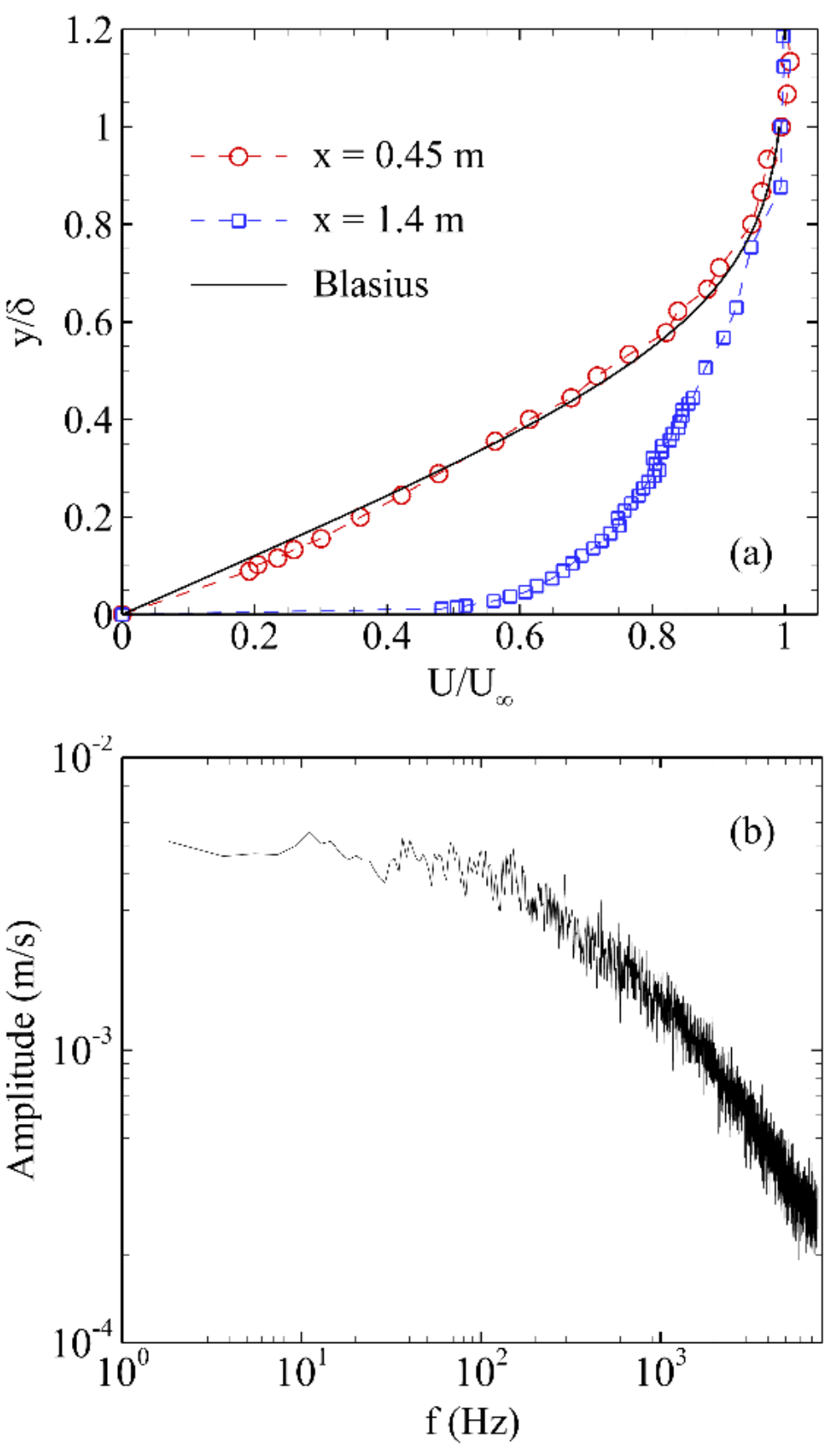

**Fig. 4 a)** Mean velocity profile in the boundary layer at $x = 0.45$ m & $x = 1.4$ m, **b)** energy spectrum of the streamwise velocity fluctuations at $y = 3$ mm ($y/\delta \approx 0.2$) inside the boundary layer at $x = 1.4$ m without suction. The dashed line in **b** indicates the -5/3 power law. $U_\infty = 26$ m/s

## 2.4 Measurements

| Suction Array | Hole diameter, $d$ (mm) | $x_{\mathrm{start}}$ (mm) | $l_{\mathrm{suction}}$ (mm) | Number of rows | Number of holes in each row | Spanwise pitch, $p_s$ (mm) |
|---|---|---|---|---|---|---|
| C4 | 0.7 | 415 | 9 | 4 | 154 | 3.5 (= $5d$) |
| C10 | 0.7 | 415 | 27 | 10 | 154 | 3.5 (= $5d$) |

Table 1 Details of the suction configurations. The streamwise pitch was $0.87p$, where $p$ is the spanwise pitch. The depth of the perforations, i.e., the thickness of the suction panel, was $l_d = 5$ mm ($l_d/d$ =7.1). $x_{start}$ is the streamwise location of the first row of perforations in the array, while $l_{suction}$ is the streamwise extent of the active part of the suction array.

Measurements were performed typically at five streamwise locations ($x = 0.45$ m, $x = 0.54$ m, $x = 0.8$ m, $x = 1.2$ m, and $x = 1.4$ m), acquiring signals from the hot-wire anemometer, the flow meter/totalizer and the reference pitot probe at a sampling rate of 15 kHz for 5 seconds (75,000 data points) at a constant freestream velocity. Two types of velocity measurements were conducted. The first was the boundary layer profile measurement acquiring velocity data at a sufficient number of wall-normal locations in the boundary layer at the sampling rate, and for the duration, mentioned above. In the second type of measurement, the hot-wire probe was located at the required $x$ and $y$ ($y$ = 1 or 1.5 mm, depending on the thickness of the boundary layer, and measured using a vernier), and velocity was recorded for a standard set of ten suction flow rates. Measurements were performed for two suction configurations: four rows of holes open in array A1 (configuration C4), and ten rows open in array A1 (configuration C10); see Table 1. Three values of $U_\infty$ were chosen: $U_\infty$ = 26 m/s, 31 m/s, and 34 m/s, corresponding to $Re_{\delta^*}$ at suction of approximately 1400, 1540, and 1600, respectively. Under optimum suction (data follows), the lowest $U_\infty$ resulted in a laminar boundary layer at the last measurement location $x = 1.4$ m, while the highest $U_\infty$ resulted in a fully turbulent boundary layer at the same location. The boundary layer displacement thickness at $x = 0.45$ m (close to $x_{\text{start}} = 0.415$ m, i.e., the first row of perforations, see Table 1) without suction was 0.91 mm, 0.81 mm, and 0.79 mm, respectively, for these three freestream velocities providing $d/\delta^* \approx 0.77$, 0.86, and 0.89 at suction. For presenting results in the following sections, the suction strength is expressed both in terms of the total suction volume flow rate per unit span, $q_s$, through an entire array of suction holes and in terms of the suction velocity, $v_s$, through each suction hole. Note that $q_s S = v_s(N\pi d^2/4)$ where $N$ is the number of holes in the suction array and $S$ is the spanwise extent of the array. The intermittency, $\gamma$, which indicates the fraction of the time the flow is turbulent at any location, is calculated by comparing to a threshold the normalized square of the second derivative of the velocity signal with respect to time. This method is similar to the one reported by Jahanmiri et al. (1997). However, the present work uses a graphical technique to choose the threshold, whereas Jahanmiri et al. (1997) use a frequency distribution technique to determine the same. An indicator function, $I$, at any time $t$ is defined to be 1 if

$$\langle \text{ND2} \rangle_{\text{m}} > \text{threshold},$$

and zero otherwise, where

$$\text{ND2} = \frac{(d^2u/dt^2)^2}{\max[(d^2u/dt^2)^2]}$$

is the normalized square of $d^2u/dt^2$ and $\langle\ \rangle_m$ denotes an $m$-point moving average; here, $m$ =15. The intermittency $\gamma$ over a desired time interval is the average of $I$ over the time interval. The threshold is selected visually to extract the turbulent parts of the signal. A threshold between 0.004 and 0.0045 has been used for the present data set. Sample variations with time of $u$, ND2, and $I$ are shown in fig. 5.

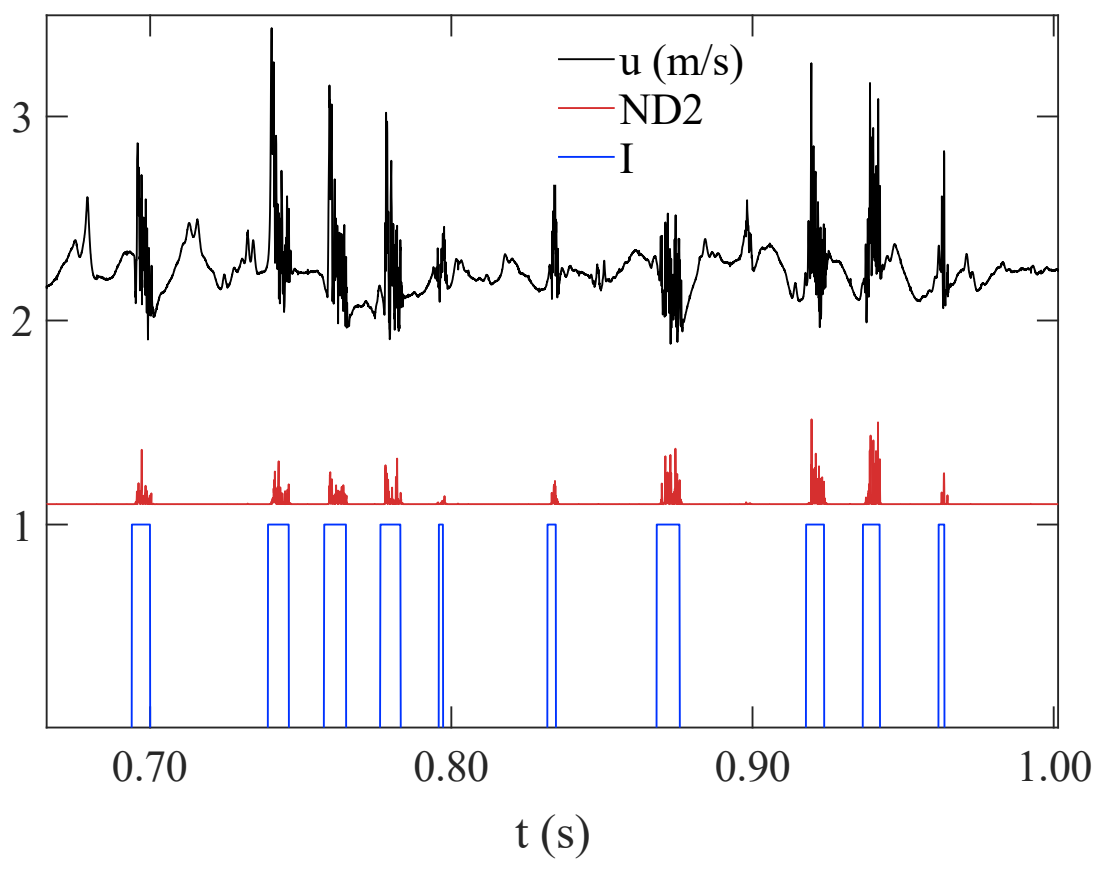

**Fig. 5** Variations with time of $u$, ND2, and the indicator function $I$ at $x = 0.54$ m, $y = 1$ mm and $U_\infty$ = 31 m/s without suction. A threshold value of 0.0045 is used for this signal to obtain intermittency of $\gamma = 0.104$. Note that the signals for $u$ and ND2 have been displaced vertically.

## 3 Results

### 3.1 Effect of suction on velocity fluctuations

The effect of suction on the boundary layer for configuration C4 and $U_\infty = 26$ m/s is presented in Fig. 6 for location $x = 0.45$ m, which is just 26 mm downstream of the last port used for suction in C4 (see Table 1). Fig. 6 a shows samples of the time series data of the instantaneous velocity for various $v_s/U_\infty$, while Fig. 6 b shows the variation of $u_{rms}/U_\infty$ with $v_s/U_\infty$. The data for the flow without suction is at the bottom in Fig.6 a and the subsequent plots for increasing suction velocity are sequentially displaced upward matching with their values of $v_s/U_\infty$ shown as the absicca in Fig 6 b. The amplitude spectra of the velocity fluctuations for selected suction velocities are presented in Fig. 6 c; the colour of each spectrum is the same as that of the time-series plot in Fig. 6 a for the respective suction strength. The time series data in Fig. 6 a show that the flow is laminar with mild waviness for all the suction strengths used. Fig. 6 b shows that at the lowest suction strength of $v_s/U_\infty = 0.07$, the velocity fluctuations ($u_{rms}$) increase slightly compared to the flow without suction, possibly indicating that the suction has caused a mild disturbance to the flow. However, on increasing the suction velocity further, $u_{rms}$ decreases continuously, attaining at the maximum suction rate of $v_s/U_\infty = 0.71$ a value approximately one-third of its value without suction . The slight waviness in the velocity signal observed for low suction velocities also gradually decreases as the suction strength is increased. Fig. 6 c shows that the amplitude of the spectra decreases with increasing frequency, falling by about three decades for frequencies up to 1000 Hz before flattening out. It is inferred that the peak in the range 46-59 Hz observed in the spectra and in the time series data was not created by induction from the electrical power with frequency of 50 Hz as it disappeared outside the boundary layer.

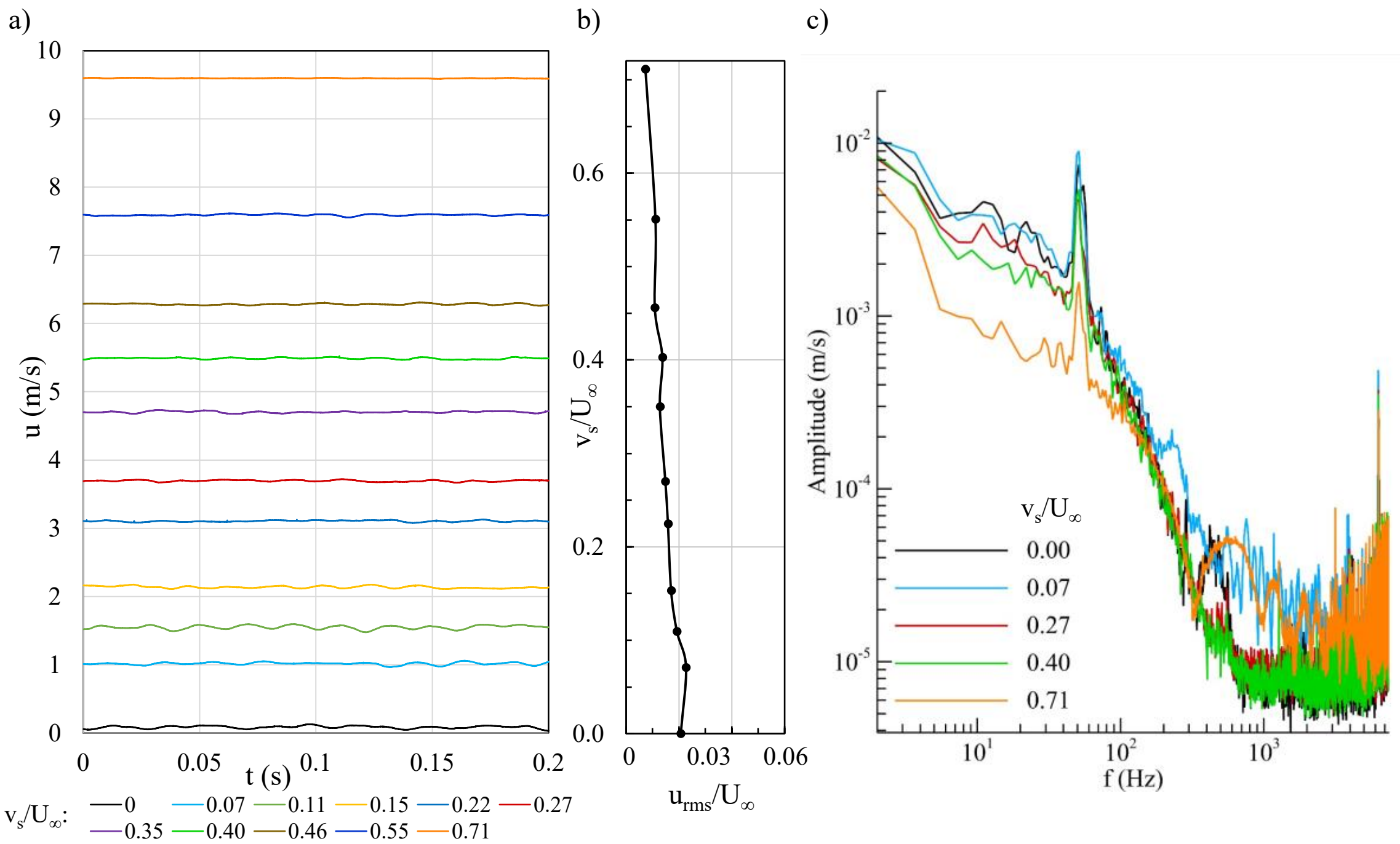

**Fig. 6** Hot-wire data at $x = 0.45$ m, $y = 1$ mm and $U_\infty = 26$ m/s for configuration C4: Variation of the instantaneous velocity with time for different suction strengths (**a**), variation of the RMS of the velocity fluctuations, $u_{rms}$, with the suction velocity (**b**), and amplitude spectra of the streamwise velocity fluctuations for different suction strengths (**c**).

A broadband hump between 300 Hz and 600 Hz is also observed in the spectra for the flow without suction. For the Reynolds number of the boundary layer at $x = 0.45$ m, $Re_{\delta^*} = 1470$, the most amplified frequency of the Tollmien-Schlichting (TS) waves can be determined from the neutral stability curve (Schlichting and Gersten 2000; Fransson and Alfredsson 2003) to be $2\pi f \nu / U_\infty^2 \approx 55$, i.e., 376 Hz, where $f$ is the frequency in Hz. Note that the TS oscillations measured at $x = 0.45$ m are a result of the amplification upstream of $x = 0.45$ m. Since the most amplified frequency is higher at lower $Re_{\delta^*}$, and hence at smaller $x$, the observed hump in the spectrum centered around 450 Hz ($2\pi f \nu / U_\infty^2 \approx 65$) may represent the TS waves. An enlarged view of the spectra is presented using a linear scale in fig. 7 for selected suction rates. It is clear from the figure that TS waves present in the flow without suction are attenuated to a significant extent even for a low suction strength of $v_s/U_\infty = 0.27$.

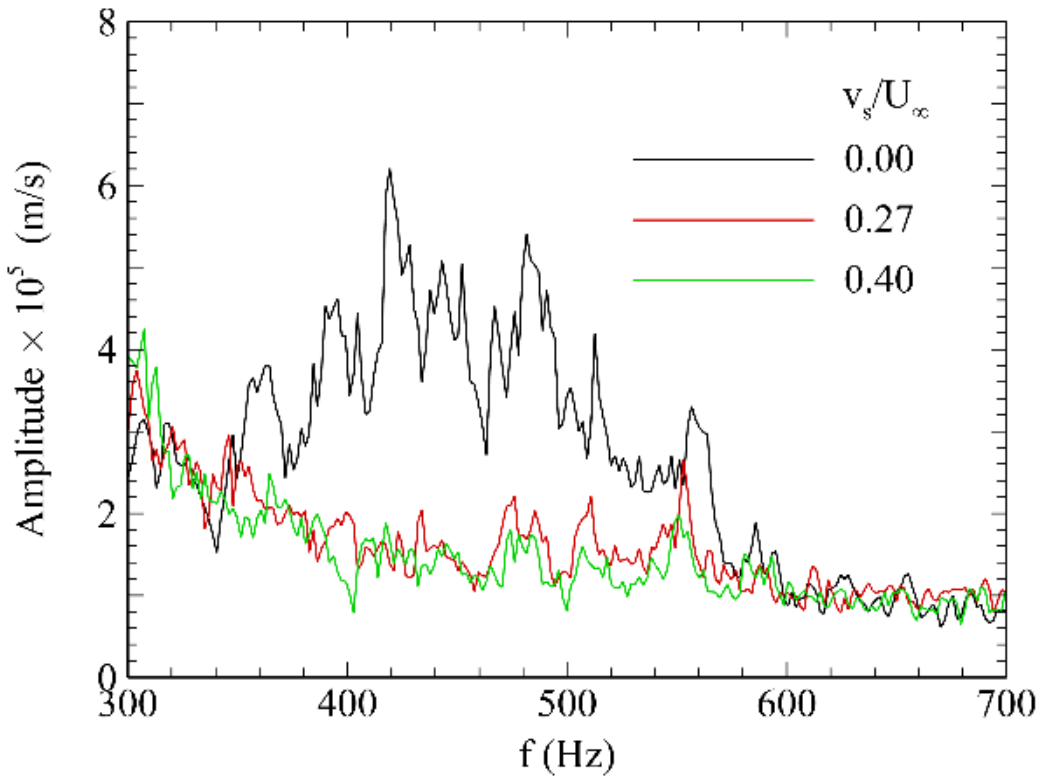

**Fig. 7** Enlarged view of the spectra using linear scales showing the attenuation of the TS waves due to suction. Data at $x = 0.45$ m, $y = 1$ mm and $U_\infty = 26$ m/s for configuration C4

The results for the most downstream location of $x = 1.4$ m are shown in Fig. 8 in a manner akin to Fig. 6, with an additional plot of the intermittency $\gamma$ in Fig. 8 b. In the absence of suction, all the parameters plotted show the characteristics of a typical turbulent boundary layer, i.e., large velocity fluctuations in the time series data, unit intermittency, and signifcant amplitudes at high frequencies. Introduction of a mild suction of up to $v_s/U_\infty = 0.11$ did not show any changes in $u_{rms}$, $\gamma$ or the spectra (hence the corresponsing spectra are not shown in Fig.8 d to maintain clarity). When $v_s/U_\infty$ is increased to 0.155, Fig.8 b shows that the intermittency decreases to $\gamma = 0.92$ and intervals of laminar flow can be observed in the turbulent velocity signal (Fig. 8 a). The intermittent nature of the flow results in higher $u_{rms}$ and higher amplitudes of the spectra (not shown) at low frequencies. At $v_s/U_\infty = 0.22$, the intermittency reduces further to $\gamma \approx 0.19$ and we can observe distinct turbulent spots in a predominantly laminar flow. The spectrum at $v_s/U_\infty = 0.27$ shows a further increase in the amplitude at lower frequencies and a reduction at all frequencies beyond about 30 Hz. When suction is increased to the maximum value of $v_s/U_\infty = 0.71$, the flow becomes fully turbulent again and the spectrum matches that for the initial turbulent flow without suction. For the intermediate suction range of $0.35 \leq v_s/U_\infty \leq 0.55$ the intermittency is zero, implying that the flow has continued to be fully laminar up to $x = 1.4$ m. The spectra show that amplitudes at higher frequencies attain a minimum at $v_s/U_\infty = 0.4$ indicating that this suction strength has the potential to extend the laminar region the farthest. Thus, $v_s/U_\infty = 0.4$ could be considered to be the optimum suction strength for this flow.

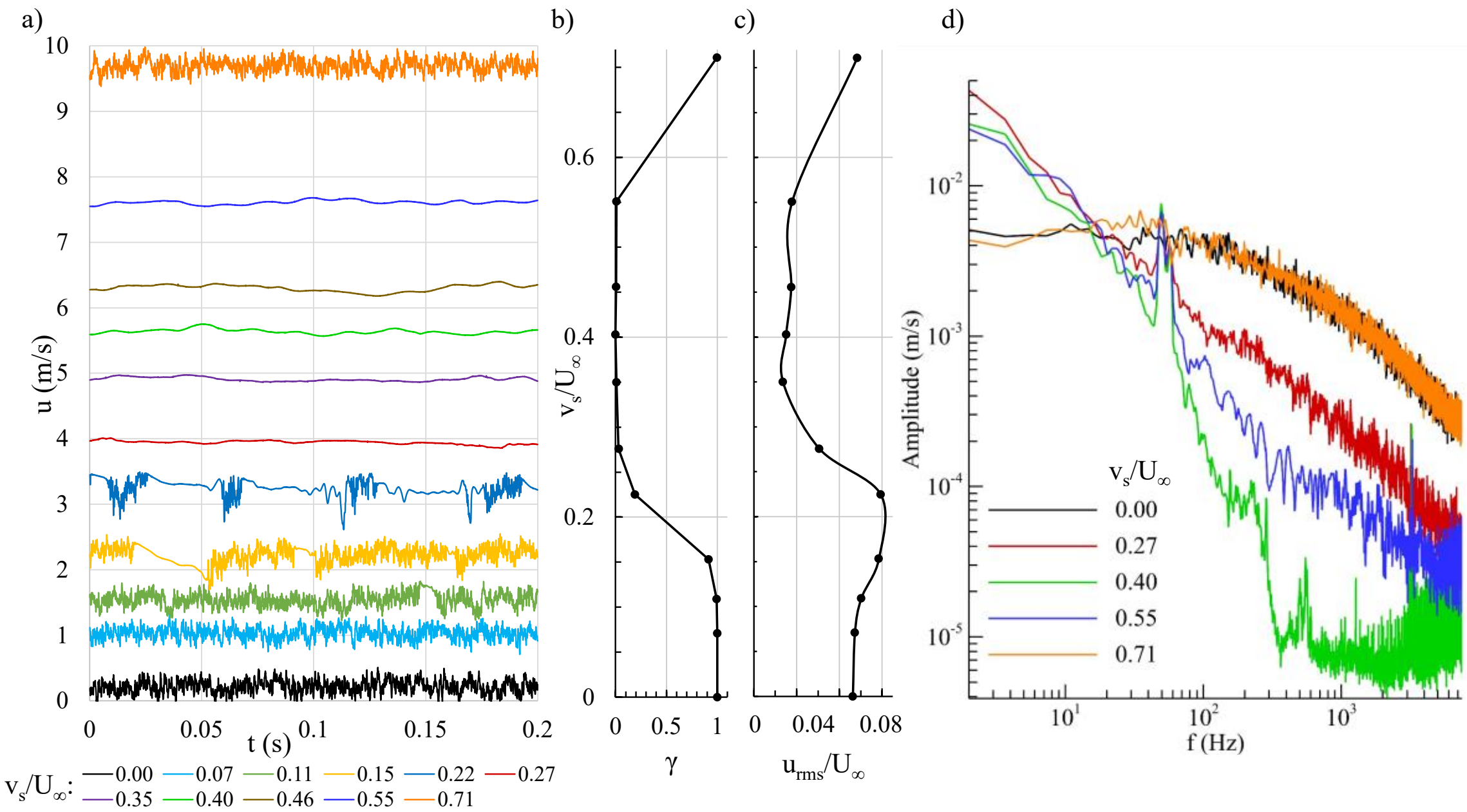

**Fig. 8** Data at $x = 1.4$ m, $y = 1.5$ mm and $U_\infty$ = 26 m/s for configuration C4. **a** Variation of the instantaneous velocity with time for different suction strengths, variation of the **b** intermittency factor, $\gamma$, and **c** $u_{rms}$ with the suction velocity, and **d** amplitude spectra of the streamwise velocity fluctuations for different suction strengths.

Having observed the effect of suction at $x = 0.45$ m and $x = 1.4$ m, we now track the development of the boundary layer between the two stations for different suction strengths. Fig. 9 a and b show the variation of $\gamma$ and $u_{rms}$, respectively, with suction strength at five selected streamwise locations. In the absence of suction, the boundary layer is laminar up to $x = 0.54$ m, has an intermittency of 0.60 at $x = 0.8$ m and is fully turbulent for $x \geq 1.2$ m. Likewise, $u_{rms}$ is the lowest for the laminar flows at $x = 0.45$ m and 0.54 m, increases, as expected, for the transitional flow at $x = 0.8$ m and decreases when the flow becomes fully turbulent at $x = 1.2$ m. The introduction of a small suction of $v_s/U_\infty = 0.07$ renders the flow fully laminar at $x = 0.8$ m. As suction is increased, the flow remains laminar for up to larger distances and as observed in Fig.8 b, a suction of $v_s/U_\infty = 0.35$ is required to retain laminar flow up to $x = 1.4$ m.

Fig. 10 shows the amplitude spectra for different suction strengths, each subfigure comparing the spectra at selected locations. In the absence of suction (Fig.10 a), we can observe that there is an increase in fluctuations at $x =$ 0.54 m as compared to those at $x = 0.45$ m due to the growth of distubances in the range 100 - 1000 Hz. A small suction of $v_s/U_\infty = 0.07$ (Fig.10 b) results in a large reduction in the amplitudes in the whole frequency range at $x =$ 0.8 m; this is accompanied by a drop in intermittency as observed in Fig. 9. As suction is increased, the amplitudes of the fluctuations in the high-frequency range continue to decrease and the boundary layer remains laminar up to greater streamwise distances. Under optimum suction, i.e., $v_s/U_\infty = 0.4$, shown in Fig.10 f, the laminar flow extends to the most downstream location, $x = 1.4$ m. As suction is further increased, Fig.10 g, h show an increase in the amplitudes beginning at higher frequencies ($f \gtrsim 300$ Hz) as the transition front advances upstream. Crowley and Atkin (2018) had also reported that transition under excessive suction involves the amplification of the high-frequency broadband velocity fluctuations. For the highest suction strength used, $v_s/U_\infty = 0.71$ (Fig.10 i), the boundary layer becomes fully turbulent at $x = 1.2$ m.

In addition to $U_\infty = 26$ m/s, measurements were performed for the configuration C4 at two more freestream velocities, $U_\infty = 31$ m/s and 34 m/s, and for configuration C10 at all these three freestream velocities. The general trends observed were similar to those discussed in this section and are not presented here in detail. In the next section, we compare the important results for the two configurations used and various freestream velocities.

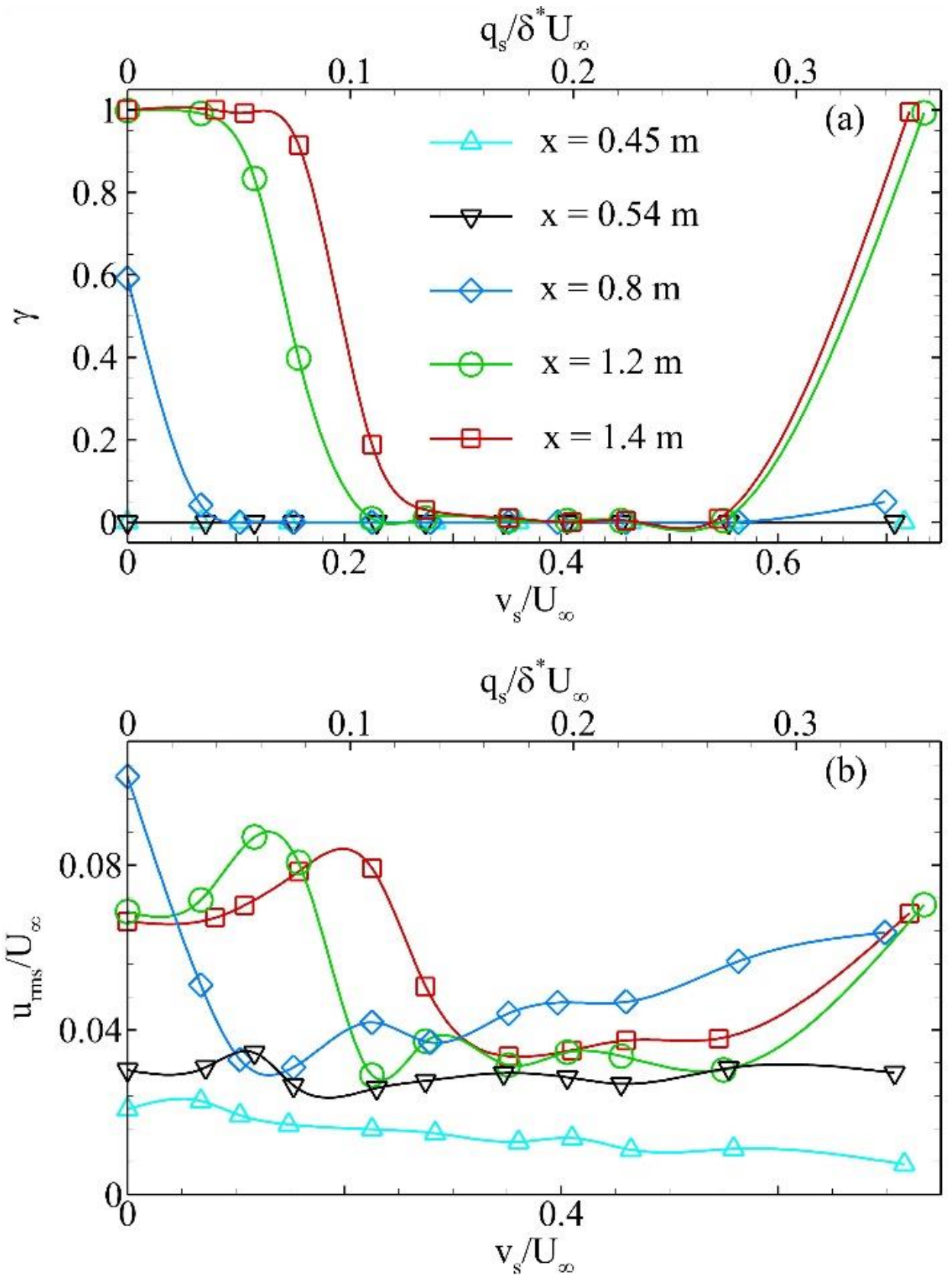

**Fig. 9** Variation of $\gamma$ (**a**) and $u_{rms}$ (**b**) with the suction strength (in terms of both suction velocity and suction volume flow rate) at various streamwise locations. $U_\infty$ = 26 m/s, configuration C4

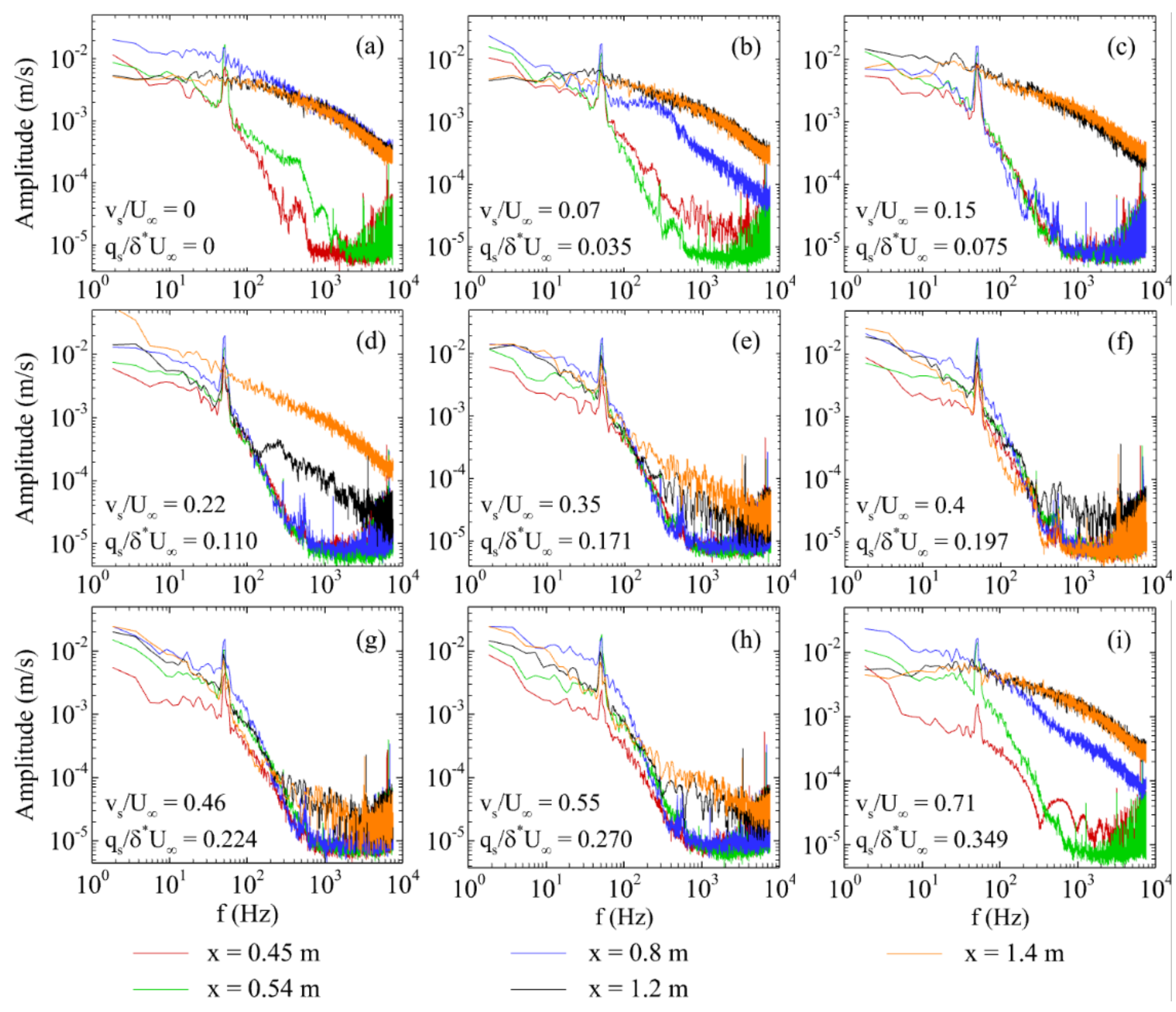

**Fig. 10** Amplitude spectra of the streamwise velocity fluctuations at various streamwise locations for selected suction strengths (expressed in terms of suction volume flow rate and suction velocity). $U_\infty$ = 26 m/s, configuration C4

### 3.2 Suction volume flow rate as the relevant parameter

The variations of intermittency with the normalized suction velocity $v_s/U_\infty$ in configurations C4 and C10 are shown in Fig. 11 a, c and e for $U_\infty = 26$ m/s, 31 m/s and 34 m/s, respectively, at selected $x$ locations (0.54 m $\leq x \leq$ 1.4 m). Fig. 11 b, d and f show the same intermittency data plotted against the normalized suction volume flow rate $q_s/\delta^* U_\infty$. When $U_\infty \geq 31$ m/s, the boundary layer did not attain a laminar state for $x > 0.54$ m irrespective of the suction strength. Under such a condition, optimum suction level is defined to be the suction strength that results in the lowest intermittency at $x = 1.4$ m. Fig. 11 shows that as suction strength ($v_s$ or $q_s$) increases from zero, intermittency decreases and reaches a minimum near optimum suction. Increasing suction beyond this value has an adverse effect, in that the transition front starts moving upstream and the intermittency at a given $x$ location starts increasing (except at locations where the flow always remains laminar). Critical suction was not encountered in this study as the transition front even for the highest suction was always downstream of that without suction.

Fig. 11 b, d, f show that when the suction strength is represented by the normalized suction volume flow rate $q_s/\delta^* U_\infty$, the value of the optimum suction strength and the general trend of the intermittency curves nearly coincide for both C4 and C10. However, the same is not true when $v_s/U_\infty$ represents the suction strength (Fig. 11 a, c, e). These comparisons show that the suction volume flow rate is a better indicator of the effects of suction than the suction velocity. So, applying suction having the same volume flow rate through either 4 or 10 rows of perforations would have a similar effect. Although previous studies have used both suction velocity and volume flow rate to quantify the strength of suction, the advantage of using the suction volume flow rate vis-à-vis suction velocity has not been brought forth.

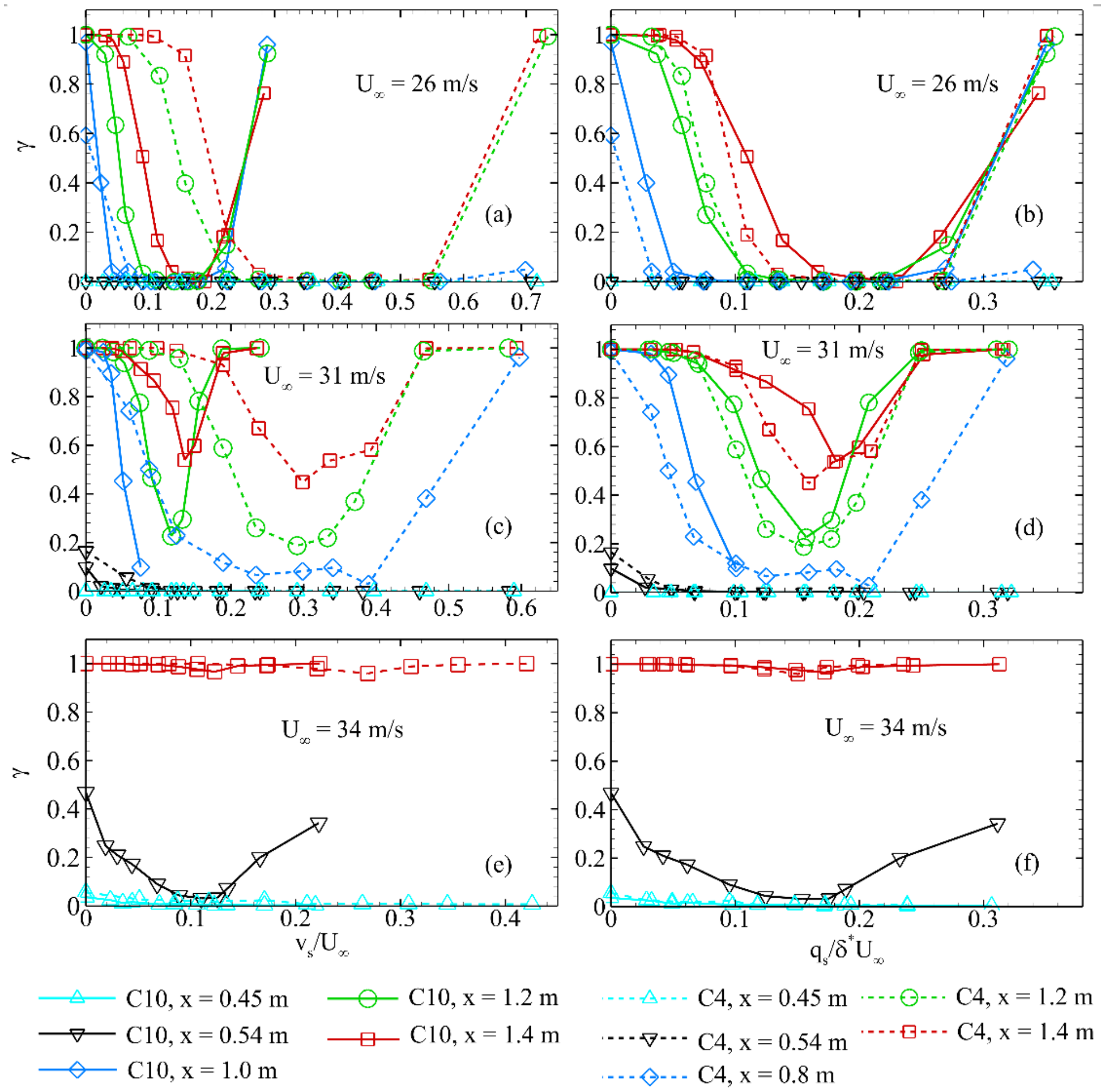

**Fig. 11** Variation of the intermittency with the normalized suction velocity $v_s/U_\infty$ (**a, c, e**), and the normalized suction volume flow rate $q_s/U_\infty\delta^*$ (**b, d, f**) at different streamwise locations for $U_\infty = 26$ m/s (**a, b**), $U_\infty = 31$ m/s (**c, d**) and $U_\infty = 34$ m/s (**e, f**). Note that data were not recorded at all streamwise locations for $U_\infty = 34$ m/s

As discussed earlier, a suction strength of $v_s/U_\infty = 0.4$ ($q_s/\delta^* U_\infty \approx 0.2$) is considered to be optimum for configuration C4 at $U_\infty = 26$ m/s, since the amplitude of the high-frequency fluctuations at the most downstream location of $x =$ 1.4 m is the smallest at this suction strength (see Fig. 8). Also, when the boundary layer was not laminar at $x > 0.54$ for any suction strength (when $U_\infty \geq 31$ m/s), we defined optimum suction to be the suction that results in the lowest intermittency at $x = 1.4$ m. Optimum values of normalized suction velocity $v_{s,\text{opt}}/U_\infty$ and suction volume flow rate $q_{s,\text{opt}}/U_\infty\delta^*$ were calculated for C4 and C10 and for all three values of $U_\infty$, and hence the Reynolds number at suction, $Re_{\delta^*}|_{\text{suction}}$, and are shown in Fig. 12. The figure shows that the blue lines (suction volume flow rate) are close to each other (i.e., nearly coinciding) while the red lines (suction velocity) are far apart, both observations again indicating that quantifying suction strength in terms of the volume flow rate is more relevant.

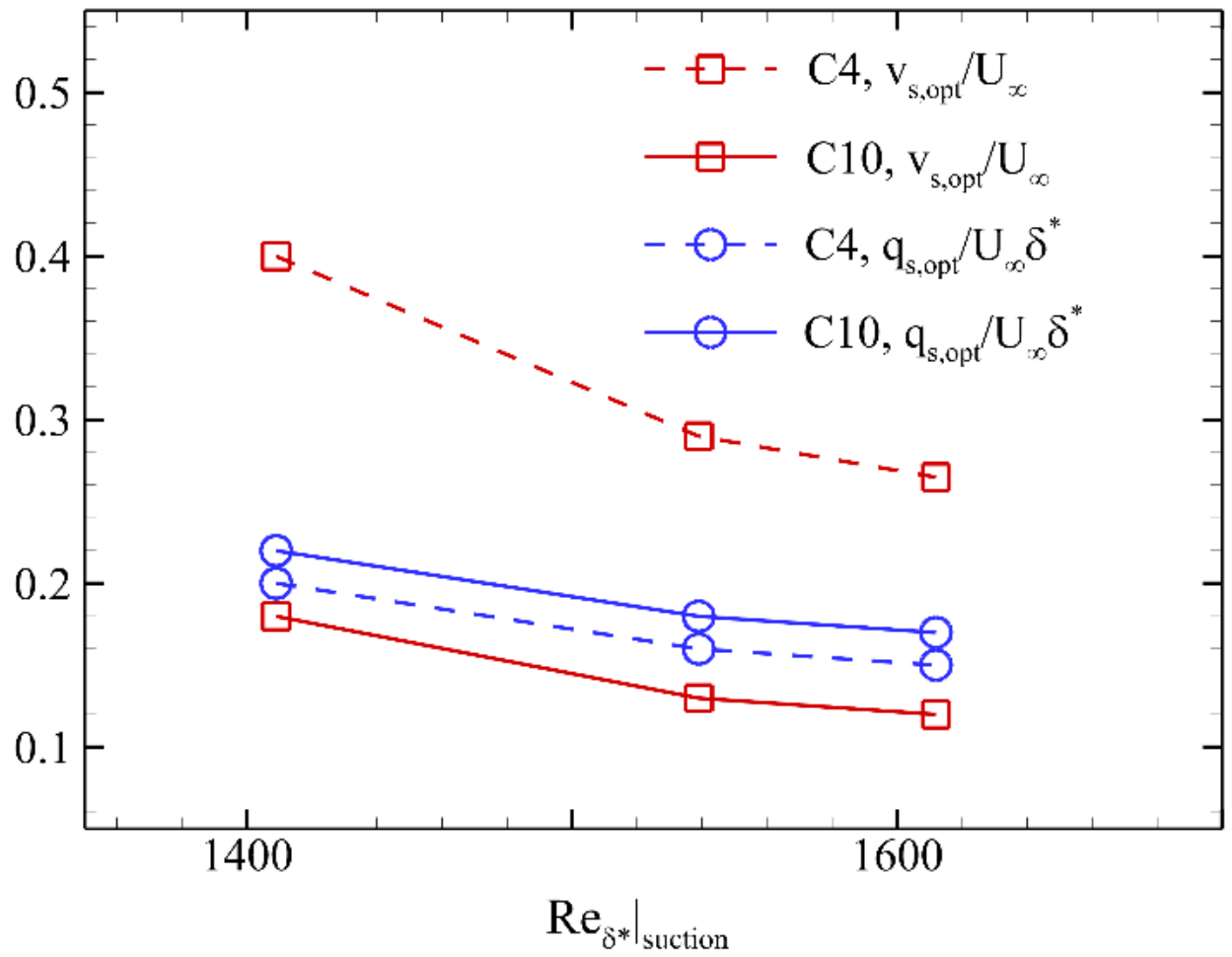

**Fig. 12** Variation of the optimum suction velocity, $v_{s,opt}/U_\infty$, and optimum suction volume flow rate, $q_{s,opt}/U_\infty\delta^*$, with $Re_{\delta^*}|_{suction}$. Cases with $Re_{\delta^*}|_{suction}$ = 1400, 1540, and 1600 correspond to $U_\infty$ = 26 m/s, 31 m/s, and 34 m/s , respectively

We now show using scaling analysis that suction volume flow rate should be a better indicator of the stabilizing effect of suction than the suction velocity. The stabilizing effect of suction has been attributed to i) a decrease in the Reynolds number of the boundary layer as the thickness of the boundary layer decreases, and ii) a decrease in the shape factor which increases the transition Reynolds number (Schlichting and Gersten 2000). For a boundary layer under a zero pressure-gradient and a suction velocity $v_s$ at the wall applied between streamwise locations $x1$ and $x2$, the momentum integral equation is

$$\frac{v_s}{U_\infty} + \frac{\tau_w}{\rho U_\infty^2} = \frac{d\theta}{dx},$$

i.e., the change in $\theta$ from $x1$ to $x2$ is

$$\Delta\theta = \int_{x1}^{x2} \frac{v_s}{U_\infty} dx + \int_{x1}^{x2} \frac{\tau_w}{\rho U_\infty^2} dx \quad \dots\dots\dots (1)$$

Thus,

$$\Delta\theta = \Delta\theta_1 + \Delta\theta_2,$$

where

$$\Delta\theta_1 = \frac{1}{U_\infty}\int_{x1}^{x2} v_s dx = -\frac{q_s}{U_\infty}$$

and

$$\Delta\theta_2 = \int_{x1}^{x2} \frac{\tau_w}{\rho U_\infty^2} dx$$

are the two contributions to $\Delta\theta$ in equation (1). Note that

$$\Delta\theta_2 = \int_{x1}^{x2} \frac{\tau_w}{\rho U_\infty^2} dx \sim \frac{1}{\rho U_\infty^2} \mu \frac{U_\infty}{\delta} \Delta x \sim \frac{\nu}{U_\infty \delta} \Delta x.$$

Thus, $\Delta\theta_2/\Delta\theta_1 \sim (U_\infty/v_s)/Re_\delta$, where $Re_\delta = U_\infty \delta/\nu$. Since $Re_\delta$ at suction is expected to be $O(10^3)$ and $v_s/U_\infty \sim O(0.1)$, $\Delta\theta_1 \gg \Delta\theta_2$, and

$$\Delta\theta \approx -\frac{q_s}{U_\infty} = \frac{v_s \Delta x}{U_\infty}.$$

Thus, the change in the momentum thickness of the boundary layer due to suction, $\Delta\theta$, and consequently, the change in the local Reynolds number $Re_\theta$ (for constant $U_\infty$), are directly related to the suction volume flow rate $q_s$ (or the product $v_s\Delta x$), and not to the suction velocity $v_s$.

### 3.3 Comparing suctions configurations

The present study has utilized two suction configurations: C4 with four rows of perforations and C10 with ten rows. Thus, the configuration C4 has shorter suction length, and hence higher suction velocity, than C10 for the same suction volume flow rate. As discussed in the previous section, the intermittency trends and the optimum suction flow rates for C4 and C10 are nearly the same. However, there exist some subtle differences between C4 and C10, which we bring out in this section.

Firstly, returning to Fig.11 b, d, it is observed that C4 has consistently lower intermittency in the optimum suction region compared to C10 in almost all cases, implying that C4 would have better prospects for delaying transition beyond $x = 1400$ mm. This observation is in agreement with the recommendation of Joslin (1998) and the observations of Cornelius (1987) that for the same suction mass flow rate narrow strips of perforation are better. It may be noted that for suction through porous panels, which is the closest to uniform suction, Reynolds and Saric (1986) observed that the streamwise length of the suction panel ($7\delta - 34\delta$) does not alter the effect of suction on the growth of the perturbations if the suction volume flow rate is constant. Thus, the present limited observations for discrete suction run contrary to the previous results for uniform suction. Secondly, Fig.12 also reveals that the optimum values of $q_s/\delta^* U_\infty$ for configuration C4 are lower than that for configuration C10. These two observations could indicate that shorter suction strips are not only better in reducing intermittency but also do it at slightly lower optimum suction value than longer suction strips. Note that the near constancy of optimum suction volume flow rate for the two configurations implies a lower optimum suction velocity for C10 (see Fig. 11 a, c, e and Fig 12). This trend for the optimum suction velocity is in qualitative agreement with the observations of Goldsmith (1957) and Blanchard et al. (1991), who reported a decrease in the critical suction velocity with an increase in the number of rows of perforations.

### 3.4 Effect of the Reynolds number at suction

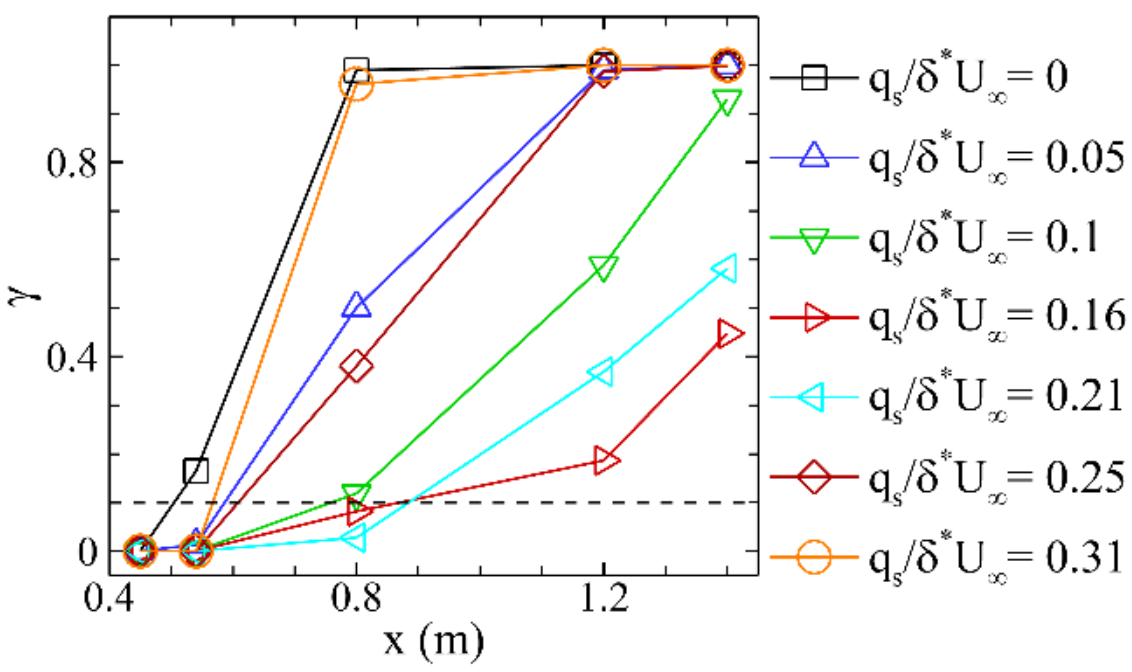

**Fig. 13** Variation of the intermittency with the streamwise distance for selected suction flow rates for configuration C4, $U_\infty = 31$ m/s. The horizontal dashed line represents $\gamma = 0.1$.

The effect of boundary layer Reynolds number at suction on the extent of the delay in the onset of transition is discussed in this section. Figure 13 shows the variation of intermittency with $x$ for various suction volume flow rates for configuration C4 at $U_\infty = 31$ m/s. For any suction volume flow rate, the $x$ location at which $\gamma = 0.1$ was estimated using linear interpolation. The onset of transition is deemed to occur at these locations, designated as $x_{\text{onset}}$, with the corresponding Reynods number $Re_{x,\text{onset}} = x_{\text{onset}} U_\infty / \nu$. The variation of $Re_{x,\text{onset}}$ with suction volume flow rate $q_s/\delta^* U_\infty$ is shown in Fig. 14 for three values of the displacement thickness Reynolds number at suction: $Re_{\delta^*}|_{\text{suction}} =$ 1400, 1540, and 1600 (corresponding to $U_\infty = 26$ m/s, 31 m/s, and 34 m/s, respectively). In the absence of suction, $Re_{x,\text{onset}} \approx 10^6$ for all flows with a small variation (attributed to experimental uncertainty and linear interpolation). For all $Re_{\delta^*}|_{\text{suction}}$, as $q_s$ increases from zero, $Re_{x,\text{onset}}$ increases, reaches a maximum, and then decreases at higher suction volume flow rates. The plateau of $Re_{x,\text{onset}}$ at $2.3 \times 10^6$ for $Re_{\delta^*}|_{\text{suction}} = 1400$ corresponds to $x_{\text{onset}} = 1.4$ m and indicates that the onset of transition occurs downstream of the largest value of $x$ for the present measurements (discussed in Sec. 3.1). Thus, the maximum increase in the onset Reynolds number for $Re_{\delta^*}|_{\text{suction}} = 1400$ seems significantly greater than 130%. This is greater than that obtained by Grappadelli et al (2021), who reported a maximum delay in transition of approximately 100% by applying suction at $Re_{\delta^*}|_{\text{suction}} \approx 1400$. In their study, the transition front was represented by the midpoint of the streamwise extent of transition as deduced from infrared imaging. Their $d/\delta^* \sim O(0.1)$ was one order lower, the number of rows of perforations substantially higher (greater than 100), and their suction rate ($0.59 \lesssim q_s/\delta^* U_\infty \lesssim 1.37$) significantly higher than in the present study. Note that for both configurations C4 and C10, $Re_{x,\text{onset}}$ for any given suction rate is consistently higher for lower $Re_{\delta^*}|_{\text{suction}}$. In particular, lowering $Re_{\delta^*}|_{\text{suction}}$ by 12.5% (from 1600 to 1400) increases the maximum increase in $Re_{x,\text{onset}}$ achieved using suction by more than 350%; refer to Fig. 14.

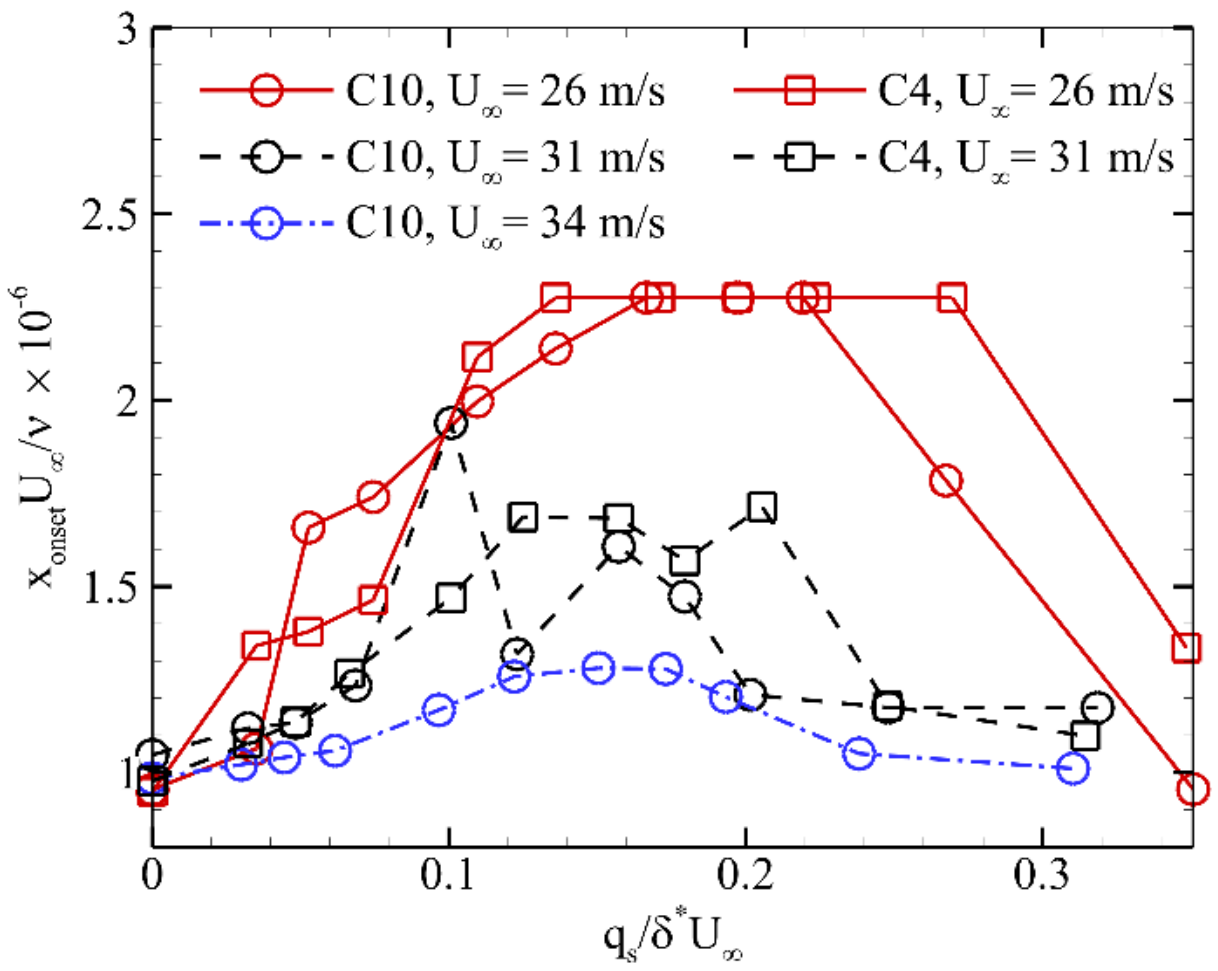

**Fig. 14** Variation of $Re_{x,onset}$ with the suction volume flow rate. Cases with $U_\infty$ = 26 m/s, 31 m/s, and 34 m/s correspond to $Re_{\delta^*}|_{suction}$ = 1400, 1540, and 1600, respectively. Data for configuration C4 at $U_\infty$ = 34 m/s were recorded only at $x$ = 0.45 m and 1.4 m. Thus, the transition locations for this case are not included due to insufficient data available for interpolation.

To characterize the state of the boundary layer subjected to suction at various $Re_{\delta^*}|_{\text{suction}}$, Fig. 15 shows the time series of the instantaneous velocity and the corresponding spectra in the boundary layer at $x = 0.45$ m in the absence of suction for different $Re_{\delta^*}|_{\text{suction}}$ (i.e., different $U_\infty$). Since suction for both configurations, C4 and C10, is applied starting from $x_{\text{start}}$ = 415 mm (see Table 1), the data at $x$ = 450 mm closely represents, with a difference in $Re_{\delta^*}$ of approxiamtely 4%, the state of the boundary layer at suction. It is clear that as $Re_{\delta^*}|_{\text{suction}}$ increases from 1400 to 1600, the boundary layer at suction changes from laminar to transitional and the intermittency of the velocity increases. Thus, suction is more effective in delaying transition when the velocity fluctuations or disturbances are smaller. This is consistent with the observation of Reed and Nayfeh (1986) based on linear stability analysis that to minimize the amplitude of disturbance of a particular wavelength at locations downstream of uniform suction, applying suction close to the lower-$Re$ branch of the neutral stability curve is better than in the region of maximum growth.

Finally, we note that the optimum suction decreases slightly as $Re_{\delta^*}|_{\text{suction}}$ increases (see Fig. 12 and Fig. 14). This trend should be expected: after the maximum delay in transition, $x_{\text{onset}}$ starts moving upstream again with an increase in $q_s$, presumably due to the destabilizing effect of the vortical structures (Gregory 1961; Meitz and Fasel 1995; MacManus and Eaton 1996, 2000) introduced by suction. This destabilizing effect may be expected to be stronger for higher boundary layer Reynolds number. A similar decrease in the critical suction velocity with an increase in the boundary layer Reynolds number at suction was reported by Gregory (1961). Thus, lowering the Reynolds number at suction from 1600 to 1400 increases the maximum delay achieved in the onset of transition, but also provides this maximum delay at greater normalized suction strength.

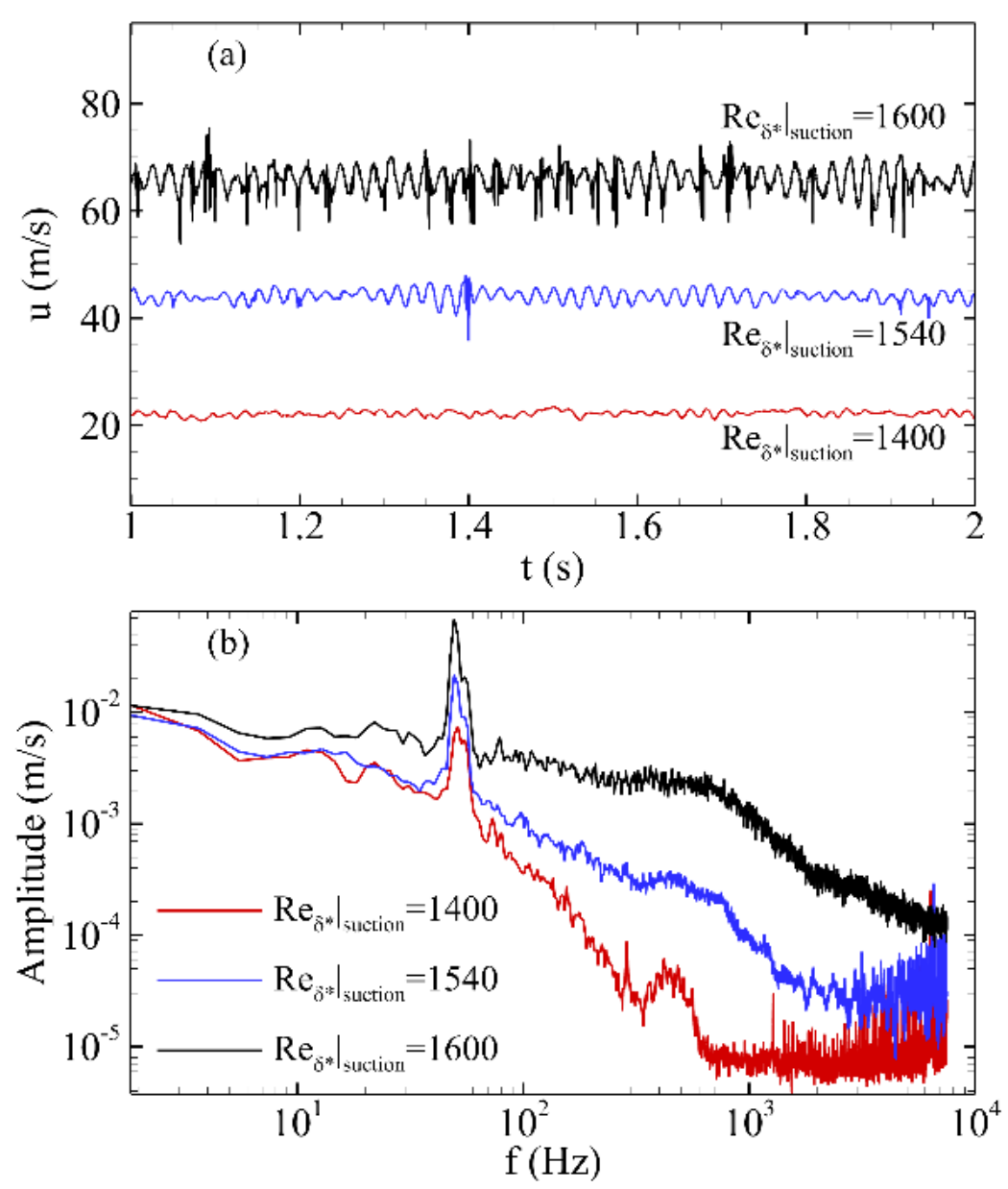

**Fig. 15** Time series of instantaneous velocity (**a**) and the corresponding spectra (**b**) in the boundary layer at $x = 0.45$ m and $y/\delta \approx 0.2$ for different $U_\infty$ (and hence, $Re_{\delta^*}|_{suction}$) in the absence of suction. Cases with $U_\infty = 26$ m/s, 31 m/s, and 34 m/s correspond to $Re_{\delta^*}|_{suction} = 1400$, 1540, and 1600, respectively

## 5 Summary

An experimental study utilizing hot-wire anemometry was performed to assess the effectiveness of suction in delaying the transition of Blasius boundary layer. Suction was applied using arrays of perforations in the form of spanwise staggered rows of holes. The Reynolds number at suction was maintained between $Re_{\delta^*} = 1400$ and 1600, while the ratio $d/\delta^*$ was maintained close to the more realistic value of 1. Two suction configurations were employed: with four and ten rows of perforations. Streamwise velocity fluctuations, intermittency and spectra were employed as diagnostics to study the effect of the suction on the boundary layer.

It was observed that even the smallest suction used in the present study, $q_s/U_\infty\delta^* \approx 0.025$, delayed the transition of the boundary layer. The applied suction decreased the amplitude of the high-frequency fluctuations and increased that of the low-frequency fluctuations. As the strength of the applied suction increased, the location of transition initially moved downstream. For a certain 'optimum' suction strength, $0.17 \leq q_{s,\mathrm{opt}}/U_\infty\delta^* \leq 0.22$, the delay in the transition obtained was the greatest. The maximum delay in the onset of transition was observed at the lowest $Re_{\delta^*}|_{\mathrm{suction}}$ investigated and was greater than 130%. As the suction strength increased beyond the optimum suction strength, the transition location moved upstream again.

The results showed that the suction volume flow rate is a better indicator than the suction velocity of the effects of suction on the boundary layer, which is also brought forth by a simple consideration of the momentum equation. Thus, irrespective of the number of rows of perforations, the effects of suction on delaying the transition are largely similar provided the suction volume flow rate is the same. In particular, for a given perforation geometry, optimum $q_s$ remains largely unchanged even if the number of holes varies. However, narrower suction strips (fewer rows of holes) do exhibit slightly lower values of optimum $q_s$ and also provide a slightly greater delay in transition at the optimum suction. Finally, it is found that applying suction at a lower Reynolds number, at least in the range $1400 \leq Re_{\delta^*} \leq 1600$, achieves a greater delay in the onset of transition but also results in slightly higher values of the normalized optimum suction flow rate.

## Statements and Declarations

**Authors' Contributions** MR contributed to the study conception, acquisition of funds, design of the experiments, data collection, data processing and analysis, and manuscript preparation, review and editing. PJ contributed to the study conception, acquisition of funds, design of the experiments, data processing and analysis, and manuscript preparation, review and editing. IS contributed to the design of the experiments, data collection, data curation, data processing and analysis, and reviewing the manuscript.

**Acknowledgements** The authors thank Mr. Manjunatha Gowda and Mr. Varadharaja for help in the experiments, and National Aerospace Laboratories (NAL) for providing the funding for this work.

**Conflict of interest** The authors declare no conflict of interest.

**Funding** The work was supported by NAL internal funding under project number: E-8-438.

**Consent for publication** All the authors of this manuscript consent to the publication of this manuscript. The necessary clearances for publication have also been obtained from NAL administration.

**Data availability and materials** The authors of this manuscript are willing to share the data upon reasonable request.